\documentclass[aps,prl,reprint,superscriptaddress,showpacs,longbibliography,amsmath,amssymb,floatfix]{revtex4-1}
\usepackage{amsmath}
\usepackage{amsfonts}
\usepackage{graphicx}
\usepackage{color}
\usepackage{hyperref}

\usepackage{bm}
\usepackage{miller}
\usepackage{siunitx}

\newcommand{\ud}{\,\mathrm{d}}

\begin{document}
\title{Distinctive Picosecond Spin Polarization Dynamics in Bulk Half-Metals}

\author{M. Battiato}
\affiliation{School of Physical and Mathematical Sciences, Physics and Applied Physics, Nanyang Technological University, 21 Nanyang Link, Singapore, Singapore}
\affiliation{Institute of Solid State Physics, Technische Universit\"{a}t Wien, Wiedner Hauptstra{\ss}e~8, 1040 Vienna, Austria}
\email{}
\author{J.~Min\'{a}r}
\affiliation{New Technologies-Research Center, University of West Bohemia, Univerzitni~8, 306\,14 Pilsen, Czech Republic}
\author{W.~Wang}%
\affiliation{Department of Physics, Biology and Chemistry, Link\"{o}ping University, 581\,83 Link\"{o}ping, Sweden}
\author{W.~Ndiaye}
\affiliation{Laboratoire de Physique des Mat\'eriaux et des Surfaces, Universit\'e de Cergy-Pontoise, 5 mail Gay-Lussac, 95031~Cergy-Pontoise, France}
\author{M.~C.~Richter}
\affiliation{Laboratoire de Physique des Mat\'eriaux et des Surfaces, Universit\'e de Cergy-Pontoise, 5 mail Gay-Lussac, 95031~Cergy-Pontoise, France}
\affiliation{DRF, IRAMIS, SPEC -- CNRS/UMR~3680, B\^{a}t.~772, L'Orme des Merisiers, CEA Saclay, 91191 Gif-sur-Yvette Cedex, France}
\author{O.~Heckmann}
\affiliation{Laboratoire de Physique des Mat\'eriaux et des Surfaces, Universit\'e de Cergy-Pontoise, 5 mail Gay-Lussac, 95031~Cergy-Pontoise, France}
\affiliation{DRF, IRAMIS, SPEC -- CNRS/UMR~3680, B\^{a}t.~772, L'Orme des Merisiers, CEA Saclay, 91191 Gif-sur-Yvette Cedex, France}
\author{J.-M.~Mariot}
\affiliation{Sorbonne Universit\'e, CNRS (UMR~7614),\\ 
Laboratoire de Chimie Physique--Mati\`ere et Rayonnement, 4 place Jussieu, 75252~Paris~Cedex~05, France}
\affiliation{Synchrotron SOLEIL, L'Orme des Merisiers, Saint-Aubin, BP~48, 91192 Gif-sur-Yvette, France}
\author{F.~Parmigiani}
\affiliation{Dipartimento di Fisica, Universit\`{a} degli Studi di Trieste, via A.~Valerio 2, 34127 Trieste, Italy}
\affiliation{Elettra-Sincrotrone Trieste S.C.p.A., Strada Statale~14, km~163.5, 34149 Basovizza, Italy}
\affiliation{International Faculty, Universit\"{a}t zu K\"{o}ln, 50937 K\"{o}ln, Germany}
\author{K.~Hricovini}
\affiliation{Laboratoire de Physique des Mat\'eriaux et des Surfaces, Universit\'e de Cergy-Pontoise, 5 mail Gay-Lussac, 95031~Cergy-Pontoise, France}
\affiliation{DRF, IRAMIS, SPEC -- CNRS/UMR~3680, B\^{a}t.~772, L'Orme des Merisiers, CEA Saclay, 91191 Gif-sur-Yvette Cedex, France}
\author{C.~Cacho}
\affiliation{Diamond Light Source, Harwell Science and Innovation Campus, Didcot~OX11~0DE, United~Kingdom}
\affiliation{Central Laser Facility, Rutherford~Appleton~Laboratory, Didcot~OX11~0QX, United~Kingdom}

\date{\today}

\begin{abstract}
Femtosecond laser excitations in half-metal (HM) compounds are theoretically predicted to induce an exotic picosecond spin dynamics. In particular, conversely to what is observed in conventional metals and semiconductors, the thermalization process in HMs leads to a long living partially thermalized configuration characterized by three Fermi--Dirac distributions for the minority, majority conduction and majority valence electrons respectively. Remarkably, these distributions have the same temperature but different chemical potentials. This unusual thermodynamics state causes a persistent non-equilibrium spin polarization only well above the Fermi energy. Femtosecond spin dynamics experiments performed on Fe$_3$O$_4$ by time-, spin-, and angle-resolved photoelectron spectroscopy confirm our model. Furthermore, the spin polarization response proves to be very robust and it can be adopted to selectively test the bulk HM character in a wide range of compounds.

\end{abstract}


\maketitle

Ultrafast magnetization dynamics covers a wide range of scientifically advanced and technologically attractive phenomena ranging from ultrafast demagnetization~\cite{Beaurepaire1996} to spin transport~\cite{Battiato2010,Rudolf2012,Battiato2012,BattiatoJAP14,Battiato16}, all-optical switching~\cite{Stanciu2007}, antiferromagnet spin dynamics~\cite{Kimel2004}, and artificial ferrimagnets~\cite{Mangin2014}. This scenario has fostered a significant effort for studying the magnetization dynamics and the distinctive interplay between the metallic and the insulating spin channels in half-metals (HMs), such as Heusler compounds (e.g., Co$_2$MnSi and Co$_2$FeSi) and oxides [e.g., CrO$_2$, {L}a$_{2/3}${S}r$_{1/3}${M}n{O}$_{3}$, and Fe$_3$O$_4$ (magnetite)] (see, e.g., Refs.~\cite{Fong2007,Venkatesan2007,Katsnelson2008} for reviews). This exotic dependence of the transport properties on the spin channels makes the physics of the HMs puzzling and striking~\cite{Felser2013,Sugahara2016}. 

Typically, HMs  show a relatively slow demagnetization, i.e., few tenths of picoseconds, which is supposed to be governed by the slow spin-lattice channel as the Elliot--Yafet spin-flip scattering is blocked in the gapped energy region~\cite{Zhang2006,Mueller2008}. However, a fast demagnetization was reported in Co$_2$Mn$_{1-x}$Fe$_x$Si and a fast spin-flip scattering path, in connection with the valence band photohole below the Fermi level~($E_{\mathrm{F}}$), was invoked to explain such a finding~\cite{Steil2010, Kaltenborn2014,Krauss2009}. Therefore, it remains unclear if and what kind of distinctive ultrafast electronic mechanisms should be expected in HMs, beyond the material dependent electron-phonon coupling. 

To address this challenge we have studied theoretically the ultrafast thermalization dynamics in HMs by solving the time-dependent Boltzmann scattering equation. A thermalization dynamics characterized by a long lasting partially equilibrated electronic distribution was identified. Remarkably, this dynamics shows peculiarities clearly discriminating the HMs from ordinary metals and semiconductors. In particular, a novel transient high energy spin polarization (SP) is found and its dynamics can be easily distinguished from other magnetization dynamics triggered by independent mechanisms such as ultrafast demagnetization~\cite{Beaurepaire1996,Eich2017}, ultrafast spin transport~\cite{Battiato2010,BattiatoJAP14,Battiato16}, or increase in magnetization~\cite{Rudolf2012,Locht15}. Notably, femtosecond spin dynamics experiments performed on Fe$_3$O$_4$ by time-, spin-, and angle-resolved photoelectron spectroscopy have successfully benchmarked the physics of our model. Finally, the experiments reported here unlock the gate for unambiguously testing the bulk half metallicity~(HMy).

A wide range of techniques has been put forward to verify HMy, however they can be made complicated for a number of reasons. For instance, the results of ferromagnetic-superconducting tunneling measurements can be strongly affected by hard-to-predict properties of the reconstructed surfaces and interfaces~\cite{Varaprasad2010}. This is also the case for Andreev reflection measurements in which the extraction of quantitative information relies on the way scattering at contacts is dealt with~\cite{Woods2004}. Transient magneto-optical Kerr effect~\cite{Mueller2008} fails since ultrafast demagnetization is not absent in all HMs \cite{Steil2010, Kaltenborn2014,Krauss2009}. Spin- and angle-resolved photoemission (PE)~\cite{Seddon2016} is in principle certainly the most direct experimental method to test HMy (see, e.g., Ref.~\cite{Jourdan2014}). Nevertheless its high surface sensitivity can raise difficulties when bulk properties are concerned. A further problem arises from the presence of polarons which strongly modifies the spectral weight at the Fermi level, hampering a direct comparison of the experimental data to state-of-the-art calculations of the PE spectra~\cite{Wang2013}. In this context identifying a clear fingerprint of bulk HMy remains an open challenge.

\begin{figure}[t]
	\centering
	\includegraphics[width=8.0cm]{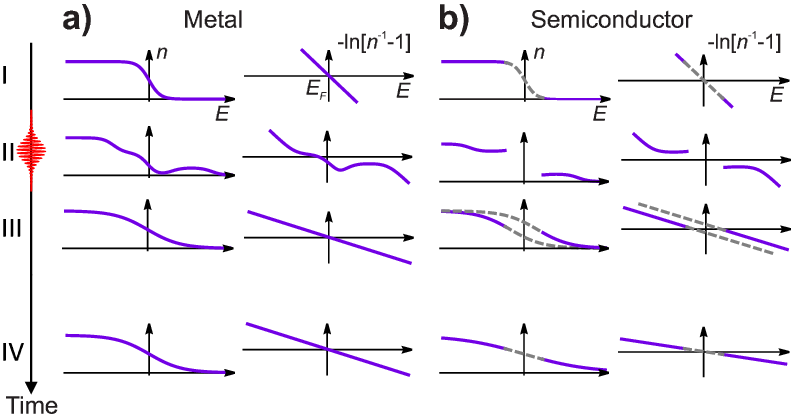}
	\caption{(Color online) Comparison of the  electron distribution function $n$ during the thermalization following a femtosecond laser excitation. (a) In a metal. (b) In a semiconductor. Both linear (left column) and logarithmic (right column) representations are plotted to highlight the location of~$E_{\mathrm{F}}$ and the electronic temperature (see text for details).}
	\label{fig:metsemi} 
\end{figure}

It is well known that electron-hole pairs created in metals under light excitation quickly decay via electron-electron scattering and form lower energy electrons and holes close to~$E_{\mathrm{F}}$, until a Fermi--Dirac (FD) distribution with a higher temperature is formed. Figure~\ref{fig:metsemi}(a) schematically illustrates the time ($t$) evolution (from top to bottom) of the electron distribution $n$ depending on the energy ($E$) in a metal after a femtosecond laser excitation. The initial FD distribution (I) evolves into a non-equilibrium distribution (II) after the laser pulse, and then, via thermalization, into another FD distribution characterized by a higher temperature~(IV). A more enlightening representation of the time dependence is given in the second column of Fig.~\ref{fig:metsemi}(a) using the $-\mathrm{ln}\left(n^{-1} - 1 \right)$ function as proposed in Ref.~\cite{Rethfeld2002}. For a FD distribution, this quantity is a linear function of $E$ with a slope proportional to $-1/k_{\mathrm{B}}T$ and a zero crossing at the chemical potential energy see Fig.~\ref{fig:metsemi}(a) panels~I, III, and IV of the second column). A non-linear behavior is associated to non-thermal distributions.

In a semiconductor, the thermalization process is completely different [Fig.~\ref{fig:metsemi}(b)]. The initial FD distribution (I) is strongly modified by the laser pulse (II) but, contrary to the metallic situation, the thermalization occurs in two steps. First electron-electron and electron-phonon scattering brings the excited electrons (hole) close to the top (bottom) of the conduction (valence) band~(III). This intermediate state is a partially equilibrated state, characterized by two FD distributions with the same temperature but different chemical potentials for the electrons and the holes [notice in (III) the different zero crossing of the two linear branches of  $-\mathrm{ln}\left(n^{-1} - 1 \right)$]. It is only on a longer timescale that a full electron-hole recombination takes place.

In a HM, where one of the spin channel is metallic and the other insulating, it is far from clear what is happening. The reason is that both channels will not behave as simply as two non-interacting populations because, even in the absence of spin-flip transitions, electrons belonging to one spin channel can scatter with the electrons of the other one.

To answer this question, we solve numerically the time-dependent Boltzmann equation for electron-electron scattering in a spin-dependent density of states (DOS) $\rho$ after a laser excitation. The time dependence of the electron distribution $n(\sigma,E,t)$ within the spin($\sigma$)- and energy($E$)-dependent DOS $\rho(\sigma,E)$ is calculated~\cite{Battiato2010, Battiato2012} as
\begin{align}
\frac{\partial n}{\partial t} = & \, S_{\mathrm{exc}}  \nonumber\\
&- \alpha\sum_{\sigma''} \int n\; \rho' (1-n') \; \rho'' n'' \; \rho''' (1-n''')\, {\ud}E'\,{\ud}E''\nonumber \\ 
&+ \alpha\sum_{\sigma''} \int (1-n)\; \rho' n' \; \rho'' n'' \; \rho'''' (1-n'''')\,{\ud}E'\,{\ud}E'' 
\end{align}
where
$n = n(\sigma,E,t)$, 
$n' = n'(\sigma,E',t)$, 
$n'' = n (\sigma'',E'',t)$, 
$n''' = n (\sigma'',E+E''-E',t)$, 
and 
$n'''' =n (\sigma'',E'+E''-E,t)$, 
and the same convention for $\rho$. The laser excitation is within the term $S_{\mathrm{exc}}=S_{\mathrm{exc}}(\sigma,E,t)$ and depends on the pump pulse parameters (duration, photon energy and intensity). Notice how electrons with different spins can scatter with each other but the total spin is preserved. Spin-flip scatterings are ignored due to their small number. We will address below how the dynamics is affected when these scatterings are included. The scattering amplitude $\alpha$ is kept as a constant parameter. The lack of energy and spin dependence of $\alpha$ is, for our purpose, an excellent approximation as the dynamics is overwhelmingly dictated by the size of the scattering phase space (i.e., the number of spin and energy conserving scatterings).

\begin{figure}[t]
	\centering
	\includegraphics[width=8.0cm]{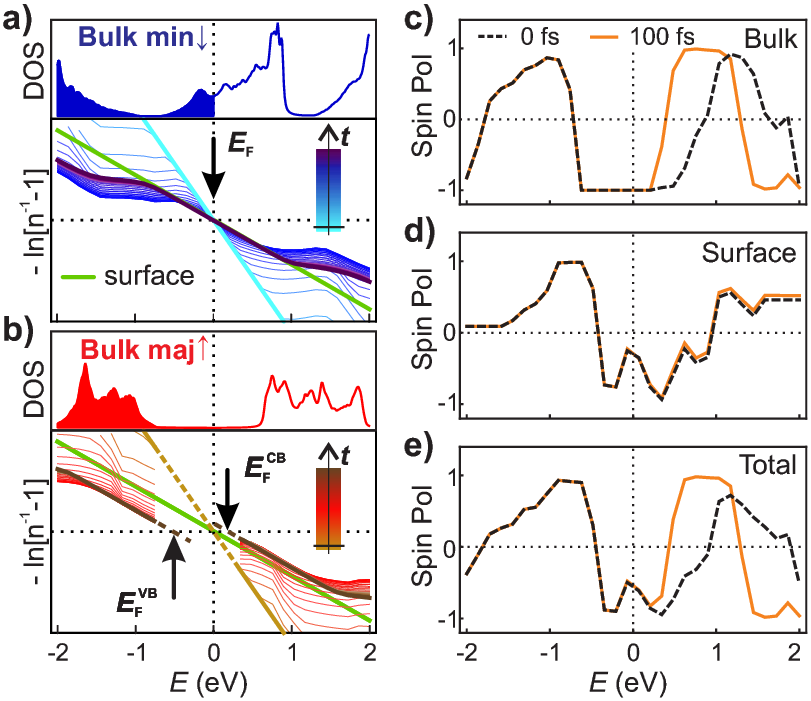}
	\caption{(Color online) Electron thermalization after a laser pulse excitation in Fe$_3$O$_4$: 
		(a) Time evolution of the electron distribution $n$ in the $-\ln\left(n^{-1} - 1 \right)$ representation for the metallic minority spin channel (bottom panel). The top panel shows the corresponding DOS of bulk Fe$_3$O$_4$.
		(b) Same for the insulating minority spin channel showing the presence of two chemical potentials. Using the metallic surface DOS, the calculations converge toward the solid green line with a single chemical potential.
		(c--e) Time dependence of the SP calculated for the bulk (HM), the surface (metal), and for a 50:50 mixture of bulk and surface contributions, respectively.}
	\label{fig:Thermalization} 
\end{figure}

For the bulk Boltzmann scattering calculations we use the Fe$_3$O$_4$ spin-resolved DOS  computed from first principles using the SPR-KKR package~\cite{Ebert2011} based on the Korringa--Kohn--Rostoker (KKR) method which uses the Green's function formalism within the multiple scattering theory. This package is based on the Dirac equation, which allows all relativistic effects to be taken into account. To treat the correlated $3d$ states of Fe, the local spin-density approximation+U (LSDA+U) method was used with the atomic limit double counting. The corresponding bulk DOS for $U = \SI{2.0}{\electronvolt}$ and $J = \SI{0.9}{\electronvolt}$ are shown in the top panels of Fig.~\ref{fig:Thermalization}(a and b); they are in very good agreement with previous theoretical results~\cite{Zang1991,Fonin2005}. The layered resolved DOS of Fe$_3$O$_4$\hkl(100) reconstructed surface from Ref.~\cite{Fonin2005} was used to take into account the (metallic) surface dynamics in solving the Boltzmann equation.

The time evolution of the electron distribution in the vicinity of $E_{\mathrm{F}}$ for the bulk minority spin (metallic channel) is presented in Fig.~\ref{fig:Thermalization}(a). Starting from a room temperature FD distribution (cyan line), the population evolves via a non-thermal distribution to a partially-thermalized distribution (from light to dark blue lines) and eventually reaches a fully thermalized population. A similar calculation performed for the metallic surface leads to a fully thermalized distribution (green line). The existence of partial thermalization  can be observed at very high and very low energies [i.e., the function $-\ln\left(n^{-1} - 1\right)$ is not a straight line]. It is due to the presence of the two regions with very low electronic densities, which act as effective band gaps. We mention this point for the sake of completeness but it is not relevant to and does not affect the conclusions of this work.

More interesting is the dynamics in the majority spin gapped channel [Fig.~\ref{fig:Thermalization}(b)]. Here the initial population evolves toward two FD-like distributions with distinct chemical potential located  (see the black arrows at zero crossing) in the valence and conduction bands (VB and CB) and with the same electronic temperature (same slope in the logarithmic representation). This result is a direct consequence of the lack of empty final states around $E_{\mathrm{F}}$ available to allow relaxation of the CB electrons and it closely resembles the behavior of an isolated semiconducting system. Hereafter we will refer to this non-equilibrium state as the partial thermalization. Due to the slow recombination of electron-hole pairs in this channel, this partially thermalized distribution persists for a long time (dark brown line). The full thermal equilibrium is reached only on a timescale that largely exceeds the computed timescale (not shown in the figure).

Such spin-asymmetric thermalization leads to the unusual situation where the SP of the partially thermalized hot carrier is strongly different from the DOS~SP. The SP [Fig.~\ref{fig:Thermalization}(c)] at the thermal equilibrium (dashed line) is the SP of the DOS (the PE matrix element effects are ignored here). After the laser excitation and an initial partial thermalization process, the accumulation of majority carriers at the bottom of the CB leads to an increase of SP in the energy range above \SI{0.5}{\electronvolt}. Due to the very slow nature of the second part of the thermalization process this out-of-equilibrium polarization persists for a long time (black line). It is here important to notice that no change in the SP is observed around the equilibrium $E_{\mathrm{F}}$ and below.
 
The above outlined SP dynamics is uniquely characteristic of~HMy. Furthermore, it can be observed even in the presence of a metallic sample surface. This is thanks to the fact that, for such a surface, the  carriers in both spin channels quickly decay to energies close to~$E_{\mathrm{F}}$. The surface SP briefly changes during the non-thermal regime due to the spin dependent laser excitation. However, within a few tens of femtosecond, the electron-electron scattering leads to a thermal equilibrium between both spin channels with a single FD distribution and the SP (black line) returns quickly to the equilibrium SP [see the dashed line in Fig.~\ref{fig:Thermalization}(d)]. In Fig.~\ref{fig:Thermalization}(e) we plot the superimposed spin dynamics of surface and bulk (i.e., 50:50 mixture) that shows how the bulk HMy effect (i.e., SP increase at high energies) is very resilient to the presence of the metallic surface. The effect is also resilient to spin-flip processes or transport between bulk and surface that would contribute  only to reduce the survival time of the out-of-equilibrium SP, but not prevent its appearance. Other effects, like polarons, that broaden the spectrum cannot mask this effect either. 

It is also fundamental to appreciate how this peculiar spin dynamics (increase of SP only well above~$E_{\mathrm{F}}$) is qualitatively different from what one would observe in the case of other types of magnetization dynamics, where the change of SP is expected either around~$E_{\mathrm{F}}$ or at all energies \cite{Beaurepaire1996,Battiato2010,Rudolf2012,BattiatoJAP14,Locht15,Battiato16,Eich2017}. Finally this dynamics cannot be confused with the one triggered by spin-asymmetric optical excitation (observed in our surface calculations) since that one survives for just a few tens of femtosecond.

We now apply our strategy to Fe$_3$O$_4$, the HMy of which is acknowledged to be difficult to measure due to its metallic surface and polaronic broadening of the PE spectra. A time-, energy-, and spin-resolved PE experiment was carried out by combining a \SI{250}{\kilo\hertz} repetition rate Ti:sapphire laser source with a homebuilt electron time-of-flight spin analyzer~\cite{Cacho2009}. The electron spin detection was achieved with a $\mu$-Mott polarimeter. Measuring the electron arrival time at each of the four detectors allows us to detect simultaneously and efficiently the energy and the spin of the photoelectrons. The third harmonic (\SI{4.65}{\electronvolt}) of the fundamental beam was generated by frequency mixing in $\beta$-barium borate crystals. For the pump beam, 10\% of the laser source was delayed and focused on the sample, giving a maximum fluence of~\SI{0.7}{\milli\joule\per\square\centi\metre}. Both pump and probe were $p$-polarized and the overall time resolution of the experiment was~$\SI{150}{\femto\second}$. The Fe$_3$O$_4$\hkl(001) thin film ($\approx \SI{50}{\nano\meter}$ thick) was epitaxially grown on MgO\hkl(100) in an ultra-high vacuum chamber and characterized by low-energy electron diffraction (LEED), x-ray magnetic circular dichroism, and x-ray photoelectron spectroscopy. The LEED pattern showed a clear $(\sqrt{2}\times\sqrt{2})\mathrm{R}45^{\circ}$ surface reconstruction. The sample was then transferred under a flow of N$_2$ to the analysis chamber. 

\begin{figure}[t]
	\centering
	\includegraphics[width=8.0cm]{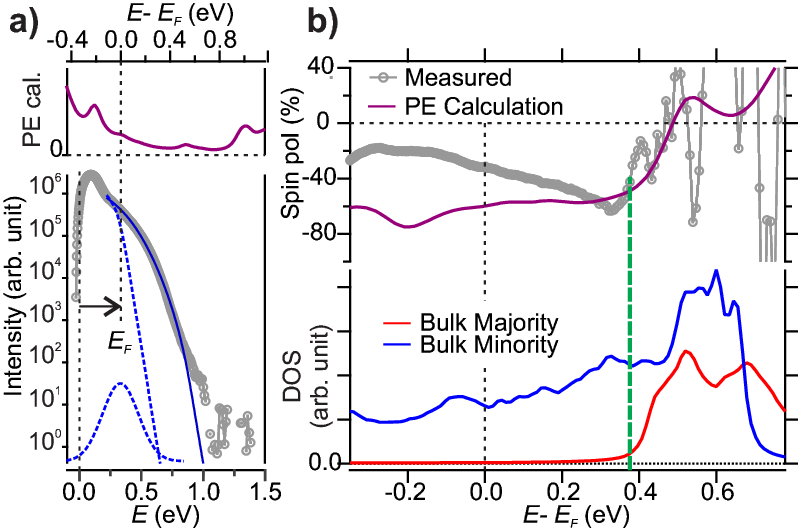}
	\caption{(Color online) Photoemission from Fe$_3$O$_4$.
		(a) The Fe $3d(t_{2g})$ EDC at \SI{4.65}{\electronvolt} photon energy (lower panel, log scale, grey dots) is fitted (solid blue line) by a function including the one-step PE calculation (top panel), a FD distribution and a broadening function (blue dashed lines).
		(b) Both measured and calculated SP (top panel) are referenced to the bulk DOS (lower panel) indicating the position (vertical green line) of the bottom of the~CB.}
	\label{fig:SpinEDCs} 
\end{figure}
 
The use of $\SI{4.65}{\electronvolt}$ photons gives access to electrons in the vicinity of $E_{\mathrm{F}}$ only, as observed on the energy distribution curve (EDC) at normal emission [Fig.~\ref{fig:SpinEDCs}(a) (lower panel, log scale, grey dots)]. The Fe $3d(t_{2g})$ spectral weight at $E_{\mathrm{F}}$ does not show any sharp edge, as it would be expected for a metal; instead it is smeared out by strong polaronic effects and initial state lifetime broadening \cite{Wang2013,Shen2007,Ihle1986}. 

The experimental EDC is very well reproduced, over three orders of magnitude (solid blue line), by the convolution of the calculated PE~\cite{Braun1996,Minar2011,Braun2014} with a \SI{330}{\milli\electronvolt} FWHM Gaussian function (dash blue lines), accounting for the polaronic effects~[\onlinecite{Wang2013}] and a FD distribution. Only the $E_{\mathrm{F}}$ position and the Gaussian width are free parameters in the fitting procedure. Due to the majority bulk band gap, the measured SP is negative around $E_{\mathrm{F}}$ and reaches only a \SI{-65}{\percent} minimum at \SI{360}{\milli\electronvolt} above $E_{\mathrm{F}}$ [Fig.~\ref{fig:SpinEDCs}(b) (top panel)] due to the metallicity of the surface. The \textit{ab initio} calculations overestimate the SP, an effect that we attribute to non trivial geometrical reconstruction of the sample surface, which is not accounted for in our calculations.

 \begin{figure}
	\centering
	\includegraphics[width=8.0cm]{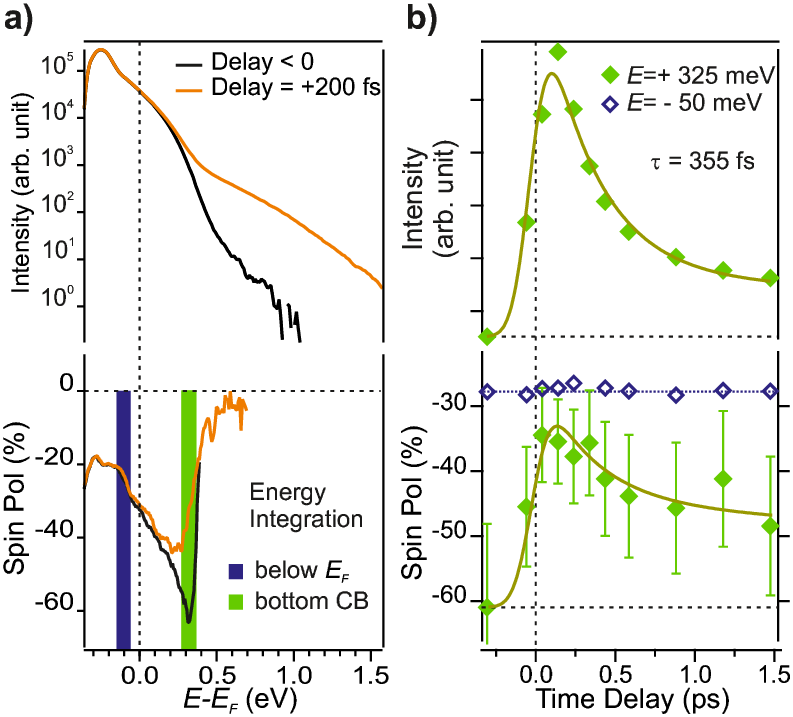}
	\caption{(Color online) Spin dynamics in Fe$_3$O$_4$ above~$E_{\mathrm{F}}$.
		(a) Spin-integrated EDC (top panel) measured before (black line) and $\SI{200}{\femto\second}$ after (orange line) pump excitation. The corresponding SP (bottom panel) increases in the energy region (green bar) at the bottom of the CB as theoretically predicted for a~HM.
		(b) The dynamics of the hot carriers (top panel) at this energy is well described by an exponential decay function characteristic of an electron-phonon cooling. The SP (bottom panel) relaxes on a similar time scale whereas the SP below~$E_{\mathrm{F}}$ (empty blue diamond), proportional to the macroscopic magnetization, remains unchanged.}
	\label{fig:SpinPol_Variation} 
\end{figure}

Next we investigate the electron spin dynamics above $E_{\mathrm{F}}$ after optical excitation. In Fig.~\ref{fig:SpinPol_Variation}(a) the EDCs (top panel) at zero delay and at $+\SI{200}{\femto\second}$ delay are compared. A clear presence of excited electrons is detected up to $+\SI{1.5}{\eV}$ above~$E_{\mathrm{F}}$. The SP below $E_{\mathrm{F}}$ (bottom panel) remains unchanged whereas, very interestingly, a clear reduction of the SP is observed in the energy region $+\SI{330}{\milli\electronvolt}$, corresponding to the bottom of the majority spin bulk~CB. The time evolution of this population [Fig.~\ref{fig:SpinPol_Variation}(b) (top panel, solid line)] is fitted by an   exponential decay function (time constant of \SI{355}{\femto\second}) plus a constant offset (to account for the extremely slow dynamics). After excitation the population relaxes quickly via a number of mechanisms such as electron-electron and  electron-phonon scattering. However a large number of carriers remain trapped at the bottom of the bulk majority spin CB, and survive at high energy for a long time ($t > \SI{1.5}{\pico\second})$. The time evolution of the SP~[Fig.~\ref{fig:SpinPol_Variation}(b) (bottom panel)] shows a similar behavior and has been fitted with the same time relaxation as for the intensity. During and shortly after the laser excitation a non trivial SP dynamics is activated. Interestingly after $\SI{1.5}{\pico\second}$ the SP does not relax back to the equilibrium value, as one would expect if the electronic system was fully thermalized. It remains $\approx25\%$ higher than its equilibrium value which is a clear evidence of the presence of a non fully thermalized population persisting in time. As this time delay is much longer than the typical electron-electron scattering time we can safely attribute this population to the partially thermalized state described in our model where a transient chemical potential is localized only in the majority spin~CB. In order to assess any possible ultra-fast demagnetization contribution in our SP dynamics we present the SP measured below $E_{\mathrm{F}}$~[Fig.~\ref{fig:SpinPol_Variation}(b) (empty diamond)]. The fact that it remains constant over time confirms that the magnetite retains its full magnetization during our time window.

The above dynamics is in striking qualitative agreement with the theoretical picture that we have presented. The accumulation of carriers at a finite energy above the Fermi energy shows the presence of a bandgap, while the transient spin polarization allows for the identification of the spin channel where the gap is located. This is in spite of the fact that a number of details are unknown, like structural and chemical reconstruction of the surface. This proves that a non-equilibrium SP at high energy, persisting well after any short-lived SP dynamics close to $E_{\mathrm{F}}$ has ended, is a reliable and resilient fingerprint of~HMy. 

In summary, we have modeled the electron thermalization following a femtosecond laser pulse in Fe$_3$O$_4$ by solving the time dependent Boltzmann scattering equation. The bulk population dynamics is found to be characterized by the formation of a persisting partial thermalization, where majority carriers reman trapped at the bottom of the conduction band. The long lasting out-of-equilibrium distribution in the CB lead to an increase of the SP well above~$E_{\mathrm{F}}$ despite the metallic nature of the surface. The agreement between experiments and theory shows that this peculiar fingerprint in the picosecond SP dynamics can be used to probe bulk HMy. This approach makes pump-probe spin-resolved PE a powerful tool to identify the HMy of materials.

\begin{acknowledgments}
The electron ToF-Spin analyzer was funded by a UK EPSRC grant (GR/M50447) and supported by STFC, and the FERMI FEL project (Elettra). M.B.~gratefully acknowledges the Austrian Science Fund (FWF) through  Lise Meitner position M1925-N28 and Nanyang Technological University, NAP-SUG for the funding of this research. J.M.~was supported by the project CEDAMNF, reg.~no.~CZ~02.1.01/0.0/0.0/15\_003/0000358, co-funded by the ERDF. 
\end{acknowledgments}

%

\end{document}